\documentclass[letterpaper]{ptephy_v1}

\begin{document}

\title{Electromagnetic Afterglows Associated with Gamma-Ray Emission Coincident with
Binary Black Hole Merger Event GW150914}

\author{
\name{\fname{Ryo} \surname{Yamazaki}}{1,\ast}, 
\name{\fname{Katsuaki} \surname{Asano}}{2},
and 
\name{\fname{Yutaka} \surname{Ohira}}{1} 
}

\address{
\affil{1}{Department of Physics and  Mathematics, Aoyama-Gakuin University, Kanagawa 252-5258, Japan}
\affil{2}{Institute for Cosmic Ray Research, The University of Tokyo, 5-1-5 Kashiwanoha, Kashiwa, Chiba 277-8582, Japan}
\email{ryo@phys.aoyama.ac.jp}
}

\begin{abstract}
The Fermi Gamma-ray Burst Monitor 
reported the possible detection of
the gamma-ray counterpart of
a binary black hole merger event, GW150914.
We show that the gamma-ray emission is caused by a relativistic outflow with
Lorentz factor larger than 10.
Subsequently, debris outflow pushes the ambient gas to form a shock,
which is responsible for the afterglow synchrotron emission.
We find that the 1.4~GHz radio flux peaks at $\sim10^5$~sec after the burst trigger.
If the ambient matter is dense enough
with density larger than $\sim10^{-2}$~cm$^{-3}$, then
the peak radio flux is $\sim0.1$~mJy, which is detectable with radio telescopes
such as the Very Large Array.
The optical afterglow peaks earlier than the radio, and if the ambient
matter density is larger than $\sim0.1$~cm$^{-3}$, the optical flux is 
detectable with large telescopes such as the Subaru Hyper Suprime-Cam.
To reveal the currently unknown
 mechanisms of the outflow and its gamma-ray emission 
associated with the binary black hole merger event,
follow-up electromagnetic observations of afterglows 
are important.
Detection of the afterglow will localize the sky position of the gravitational wave
and the gamma-ray emissions, 
and it will support the physical association between them.
\end{abstract}


\maketitle


\section{Introduction}

On September 14, 2015 the Laser Interferometer Gravitational-Wave Observatory (LIGO)
detected a transient gravitational wave (GW) signal of a binary black hole (BH) merger
\cite{Abbott2016}.
Analysis of this event, GW150914,
showed that an energy of $\sim3M_{\odot}c^2$ went into GWs and that
the  luminosity distance to the source was $\approx410$~Mpc,
corresponding to a redshift of $z\approx0.09$
  \cite{Abbott2016,LIGO2016a}.
LIGO provided us with direct evidence for the existence of binary BH systems and
existence of their merger.
This is the beginning of GW astronomy.

It is also surprising that  the {\it Fermi} Gamma-ray Burst Monitor (GBM) 
detected a possible gamma-ray counterpart 
0.4~sec after the detection of GW150914 \cite{Connaughton2016},
while electromagnetic counterparts from binary BH mergers had hardly been expected
before this event.
The gamma-ray emission lasted for about 1~sec.
The observed fluence between 10 and 1000~keV was 2--$3\times10^{-7}$erg~cm$^{-2}$,
so that the isotropic-equivalent gamma-ray energy $E_\gamma$ is
$5\times10^{48}$~erg, which is only a small fraction ($\sim10^{-6}$)
of the energy radiated by GW.
The gamma-ray luminosity in the 1~keV--10~MeV band is measured as
$L_\gamma=1.8\times10^{49}$~erg~s$^{-1}$, which is again a very tiny fraction of
the peak GW luminosity of $200 M_{\odot}c^2$~s$^{-1}$ ($\approx 10^{-3}c^5/G$).
A possible peak energy in the spectrum is a few MeV,
and there is no evidence of photons above $\sim 5$~MeV.
The spectrum seems hard considering its small $E_\gamma$,
which makes the event an outlier of the relation for short gamma-ray bursts (GRBs) \cite{Li2016}.
So far, no other electromagnetic, including neutrino, counterpart has been reported
\cite{Abbott2016_followup,Adrian2016,Annis2016,Evans2016,FermiLAT2016,
Kasliwal2016,Santos2016,Savchenko2016,Smartt2016, Troja2016}.
This surprise already triggered theoretical works 
\cite{Li2016,Kotera2016,Loeb2016,Murase2016,Perna2016,Stone2016,Zhang2016}.
Note that the significance of the GBM emission is not high, and that the observed GBM fluence 
contradicts the INTEGRAL upper limit \cite{Savchenko2016}.
Hence, at present, the gamma-ray emission associated with binary BH merger
event has not been firmly established.

In this paper, we point out that 
if the GBM detection of gamma-rays coincident with GW150914  is real, then
the gamma-ray emission arises from 
relativistically expanding material with a bulk Lorentz factor larger than 10
(section~2).
Then, we consider the afterglow emission which is the synchrotron emission of
relativistic electrons accelerated at a shock propagating into the ambient matter
(section~3).
Finally, in section~4, we compare our model prediction with current upper limits
in the X-ray, optical, and infrared bands. 
Throughout this paper, we use the ``TT+lowP+lensing+ext''
cosmological parameters from Table~4 of \cite{Planck2015}.

\section{Gamma-ray Emission}

The horizon scale of the final BH, whose mass and spin are
$M_{\rm BH}=62 M_{\odot}$
and $a=0.7$ \cite{Abbott2016,LIGO2016a}, 
is $R_{\rm BH}=(1+\sqrt{1-a^2})G M_{\rm BH}/c^2\sim 1.6 \times 10^7$~cm.
The radius of the innermost stable circular orbit is $R_0 \sim 2 R_{\rm BH}$ for $a=0.7$.
If the energy source of the gamma-ray emission is
the gravitational energy of accreting matter,
the required mass is $M_{\rm acc}
\sim E_\gamma R_{\rm BH}/(G M_{\rm BH}) \sim 10^{-5} M_{\odot}$
for $E_\gamma\sim10^{49}$ erg.
The required mass is so tiny compared to $M_{\rm BH}$ that
it is not surprising even if a small part of the circum-binary disk
supplies such a small mass $M_{\rm acc}$ onto the BH.
Another possible mechanism for the energy release is
the dissipation of the magnetic energy $\sim B^2 R_{BH}^3/6$,
which implies $B \sim 1.2 \times 10^{14}$ G.
Alternatively, the Blandford-Znajek (BZ) process may extract
the rotation energy of the BH \cite{BZ77} and produce
Poynting-flux-dominated jets.
The required magnetic field is $B \sim \sqrt{16 L_\gamma/(0.05 \pi c a^2 R_{\rm BH}^2)}
\sim 1.7 \times 10^{13}$ G \cite{Tchekhovskoy2011} 
for $L_\gamma \sim 10^{49}$ erg~$\mbox{s}^{-1}$.

First, let us consider the fireball model \cite{SP90},
which has been applied to GRBs.
A huge energy release in the localized region around the BH leads to
an optically thick radiation-dominated plasma, whose components are photons, electrons
and positrons.
If the fireball energy is continuously injected into a volume of $\sim 4 \pi R_0^3/3$,
the energy density is estimated as $e_0=L_\gamma/(4 \pi c R_0^2)$.
In this case, the initial temperature  is
$T_0=(60\hbar^3 c^3 e_0/11 \pi^2)^{1/4} \sim 90$~keV,
which is too low compared to the observed photon energy
(here the Boltzmann constant is taken as unity).
On the other hand, if an energy $E_\gamma$ is promptly released,
the energy density $3 E_\gamma/(4 \pi R_0^3)$ corresponds to an initial
temperature of $T_0\sim 0.7$~MeV.
The fireball starts to expand and is accelerated by its own pressure;
the Lorentz factor grows with the radius as $\Gamma \propto R$
and the temperature drops as $T' \propto R^{-1}$
(the photon energy density $e' \propto T'^4 \propto R^{-4}$)
\cite{Piran1993}. The photon temperature for the observer
$T_{\rm obs}=\Gamma T'=T_0$ is constant, and consistent with the observed spectrum.
The number density of the electron--positron pairs in the complete thermal equilibrium
decreases with temperature as $n'_\pm=4 (m_{\rm e} T'/2 \pi \hbar^2)^{3/2}
\exp{(-m_{\rm e} c^2/T')}$.
In the baryon-free limit, 
the photosphere radius at which $n'_\pm \sigma_{\rm T} R/\Gamma=1$
is calculated as $R_{\rm ph}=36 R_0$
for the initial temperature $T_0=0.7$~MeV, 
which means $\Gamma=36$ at the photosphere.
The luminosity of the photospheric emission is 
$L_\gamma=4 \pi c R_0^2 \Gamma^2 e'
\simeq 3 c E_\gamma/R_0 \sim 2.9 \times 10^{52}$ erg~s$^{-1}$,
which is much higher than the observed value as a natural consequence
of adjusting the temperature rather than the luminosity.
Therefore, the baryon-free photospheric model cannot reconcile the temperature with 
the luminosity.
In addition, the emission timescale $\sim R_0/c\sim1$~msec is too short in this case.
Even for the trapped fireball model \cite{Thomp95}, in which the fireball
is trapped in the magnetosphere of the BH, the contradiction
between the temperature and luminosity remains unsolved.

Therefore, the observed gamma-ray emission is not due
to photospheric emission from the baryon-free fireball.
The fireball may be contaminated by baryons.
In such a case, the fireball expansion evolves to a relativistic outflow
of baryons, which may emit gamma-rays at the outer radius
via internal shock etc. \cite{Piran1999} or
relic thermal photons at the baryonic photosphere.
If the observed emission is coming from the baryonic photosphere,
much more energy is required as the bulk kinetic energy of the baryons.
The maximum gamma-ray energy is a few MeV in the observation.
If the bulk Lorentz factor is higher than 10,
the energies of those photons are below $m_{\rm e} c^2$
in the comoving frame of the outflow.
Then, electron-positron pair production in the source
can be avoided, so that gamma-rays can safely escape from the source.
For a baryonic fireball with the final Lorentz factor $\Gamma$
and total luminosity $L_{\rm iso} \equiv L_\gamma/f_\gamma$,
the photosphere is $R_{\rm ph}=L_\gamma \sigma_{\rm T}/(4 \pi \Gamma^3 m_{\rm p} c^3 f_\gamma)$.
To set the emission radius $R \sim 2 \Gamma^2 c \Delta t$
(the time delay $\Delta t \sim 0.4$ s between gamma-ray and GW) outside the photosphere,
$\Gamma>22$ is required.
On the other hand, the baryon amount in the Poynting-flux-dominated jet model
is not constrained well, so that 
the minimum Lorentz factor required to avoid $\gamma
\gamma$ absorption may be $\sim 10$.
Even after the gamma-ray emission, a significant amount of the kinetic energy
($\sim 10^{49}$ erg) may remain as an outflow,
which plunges into the ambient medium. Similarly to the model for  binary neutron star
merger \cite{Nakar2011,Piran2013,Takami2014},
we can expect afterglow emission from the forward shock formed by 
the outflow.

\section{Afterglows}

The bulk of gamma-ray emitting materials pushes the ambient medium  to form
a shock, generating high-energy electrons which produce synchrotron radiation.
We calculate such afterglows in the same procedure as for GRBs
\cite{Piran1999,Zhang2004,Kumar2015}.
Initially the ejecta mass $M_{\rm ej}$ 
is injected at the origin $r=0$
with expanding bulk Lorentz factor $\gamma_0$.
The burst energy is then $E_0=\gamma_0 M_{\rm ej}c^2$, with which the material expands outwards.
As will be shown later, given the energy $E_0$,
the parameter $\gamma_0$ only affects the very early stage of the afterglow.
The uncertainty in $\gamma_0$ does not disturb the prediction of
the late afterglow.
We assume spherical expansion --- jet collimation is not considered
(see section~4 for a discussion).
The ambient matter is uniform with density $n_0$.
Following the previous work of \cite{Huang2000}, the dynamics of the shock is
numerically computed.
In deriving the observed flux density of the synchrotron emission $F_\nu$ at frequency $\nu$,
we integrate the emissivity over the equal arrival time surface
\cite[e.g.,][]{Sari1998,Panaitescu1998,Granot1999,Huang2000}, assuming a 
thin shell emission approximation.
Using the standard convention in the GRB community,
the power-law index of the electron distribution $p$, together with
the microphysics parameters $\epsilon_e$ and $\epsilon_B$, determines
the emissivity profile in the comoving frame.
Here, we adopt the form of the electron distribution 
$dN/d \gamma_e\propto(\gamma_e-1)^{-p}$
as given in \cite{Huang2003}.
In calculating the radio synchrotron emission, we take into account the
effect of synchrotron self-absorption (SSA).
The optical depth $\tau_a(\nu)$ to SSA was given by, for example, 
Eq.~(52) of \cite{Panaitescu2000}.
The self-absorption frequency $\nu_a$, at which $\tau_a(\nu_a)$, is generally below
the radio band, so that the emission is optically thin.
We adopt $E_0=1\times10^{49}$~erg, $\gamma_0=10$, 
$p=2.3$, $n_0=1$~cm$^{-3}$, $\epsilon_e=0.1$, $\epsilon_B=0.01$, and $z=0.09$
as fiducial parameters.

The red curves in Figure~1 show the afterglow light curves in the 1.4~GHz radio band
for the fiducial parameters.
The flux initially increases and there is a small bump at 
$t\approx3\times10^5$~sec after the GW event.
In our case, the observed $F_\nu$ spectrum has a maximum at
$\nu_{\rm peak}={\rm max}\{\nu_a,\nu_m\}$, where
the minimum frequency $\nu_m$ is the characteristic frequency of the synchrotron emission by
electrons with the minimum Lorentz factor $\gamma_m$.
The frequency $\nu_{\rm peak}$ decreases with time, and crosses the observation 
frequency of 1.4~GHz around the observer time of the bump in the light curve.
In our case, $\nu_a$ and $\nu_m$ are comparable at that time.
The peak flux and the peak time are consistent with the result in
Morsony~{\it et~al.}~\cite{Morsony2016},
which appeared in arXiv just after the submission of this paper.
At $t=5.5\times10^5$~sec,
the shock  reaches the radius at which $\gamma\beta=1$ , where
$\beta$ is the shock velocity divided by the velocity of light and 
$\gamma=(1-\beta^2)^{-1/2}$.
After that, the expansion velocity becomes Newtonian.
Furthermore, we can see $\gamma_m\ll\gamma_a\ll\gamma_c$ if $t>3\times10^6$~sec,
where $\gamma_a$ is the electron 
Lorentz factor whose typical photon frequency corresponds to the absorption frequency, 
and $\gamma_c$ is the electron Lorentz factor above which synchrotron cooling is significant.
The radio flux monotonically decreases after the peak.
The shock dynamics is roughly approximated by the Sedov scaling law,
$r\propto t^{2/5}$.
In this case, the flux is given by $F_\nu\propto m(r)B(\nu/\nu_m)^{-(p-1)/2}$
\cite{Gao2013b},
where $B$ and $m(r)$ are the magnetic field strength in the comoving frame and
the swept-up mass of ambient matter at the shock radius $r$, respectively.
Using the non-relativistic Sedov scalings,
$B\propto\beta\propto t^{-3/5}$,  $m(r)\propto r^3\propto t^{6/5}$,
and $\gamma_m\propto\beta^2\propto t^{-6/5}$,
one can see $\nu_m\propto\gamma_m^2B\propto t^{-3}$ and
$F_\nu\propto t^{-(15p-21)/10}\nu^{-(p-1)/2}$,
which is the same as predicted for the binary neutron star merger
event \cite{Nakar2011,Piran2013,Gao2013a}.
This scaling explains the decline.

One can see the parameter dependence from Figure~1.
A change of $\gamma_0$ only affects 
the flux at very early epochs.
As seen in the left panel of Figure~1, 
the flux decays more rapidly for larger values of $p$.
The number density of the ambient matter is highly uncertain, so that
we take a wide range of $n_0$ in  the right panel of Figure~1.
The radio emission is brighter for larger $n_0$.

Figure~2 shows optical R-band light curves, similarly to Figure~1.
The peak of the optical afterglow is earlier than for radio.
The peak epoch roughly corresponds to the onset time of
deceleration of the outflow \cite{Sari1999}, which is estimated as
$t_{\rm dec}=(3E_0/32\pi\gamma_0^8n_0m_pc^5)^{1/3}=4.2\times10^3$~sec
for our fiducial parameters.
After the peak, the optical flux monotonically decreases.

\begin{figure}[t]
  \begin{center}
        \includegraphics[width=70mm]{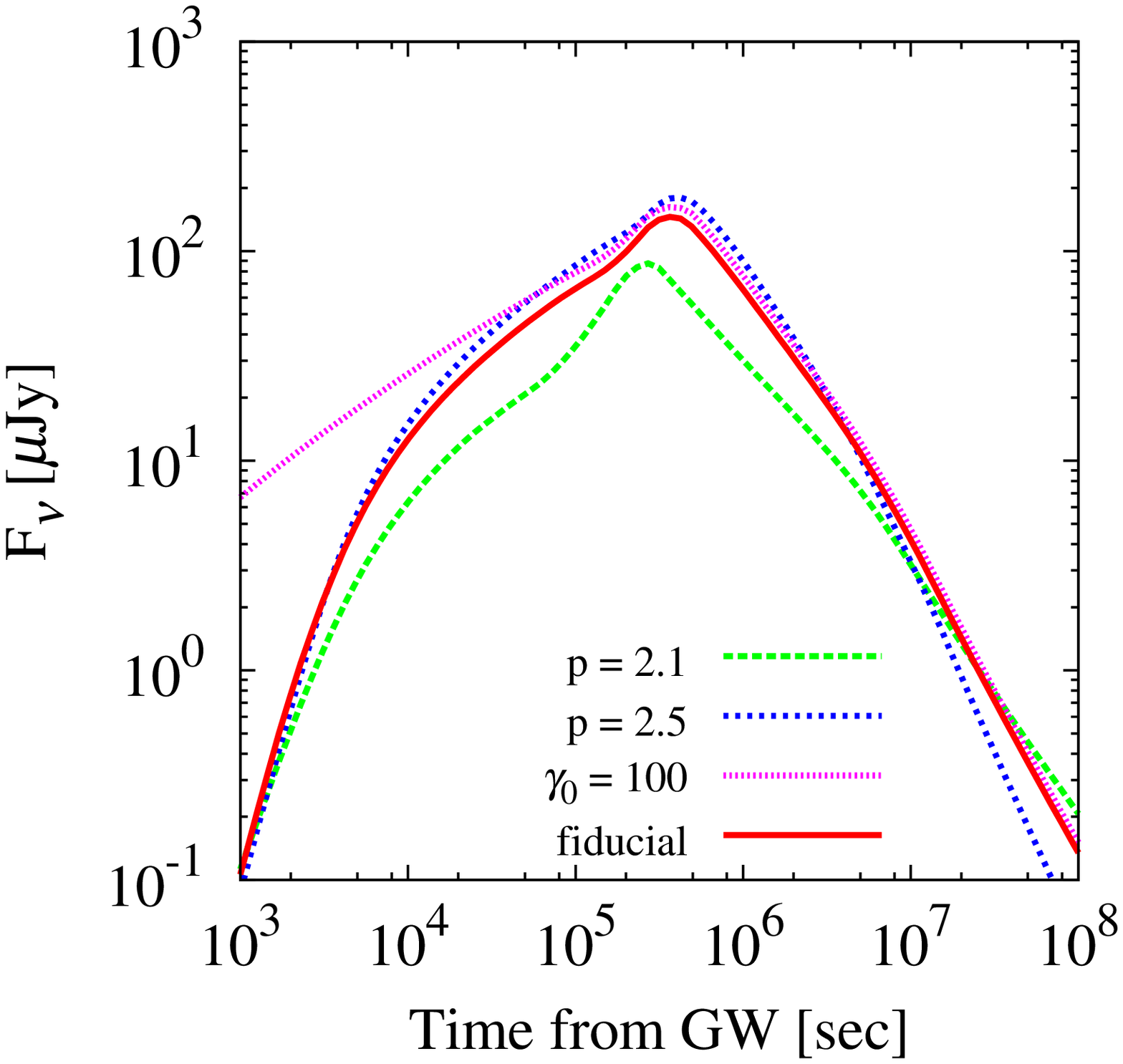}
                \includegraphics[width=70mm]{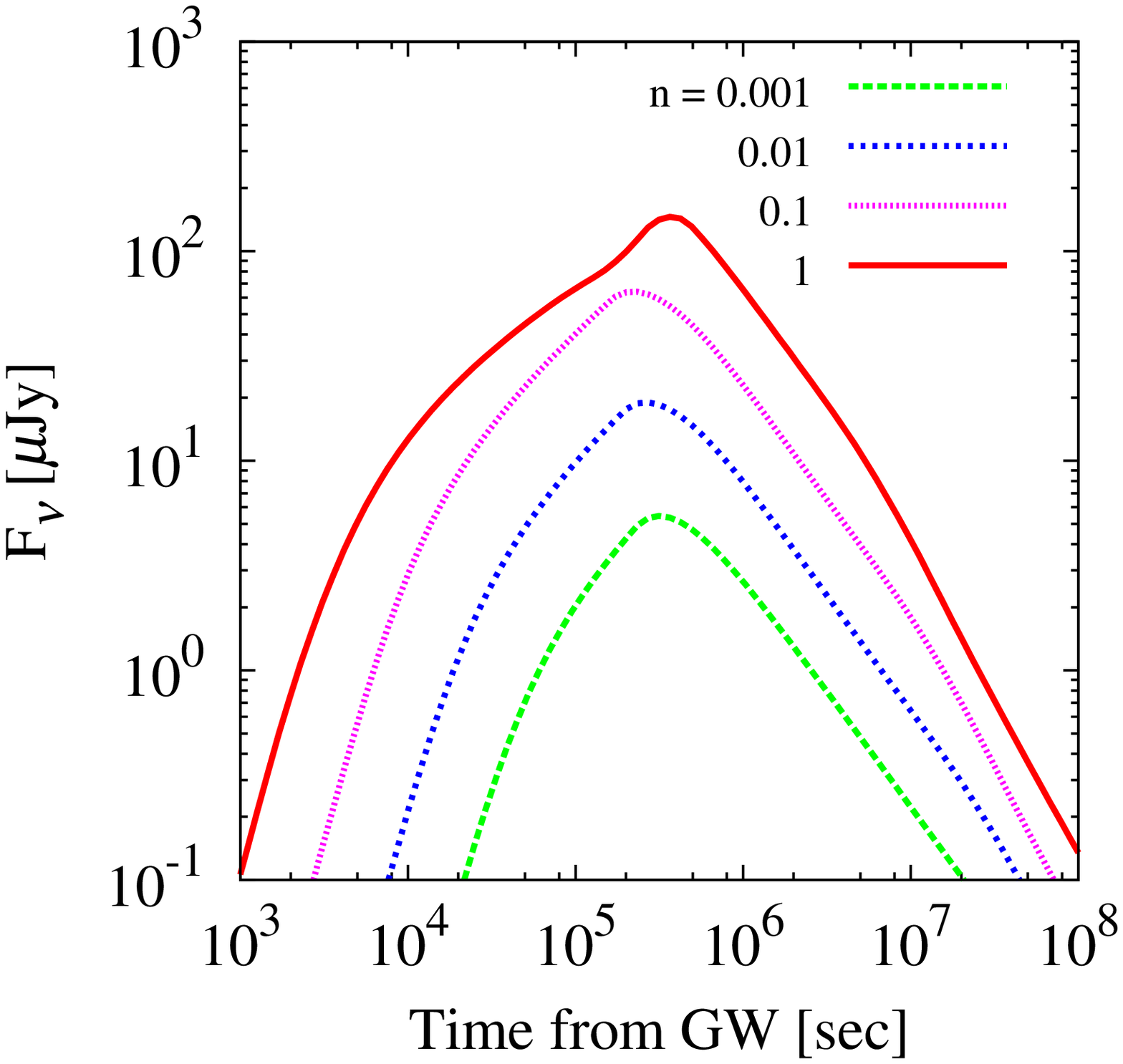} 
    \vspace{-1.3 cm}
    \caption{Radio afterglow light curves at 1.4~GHz.
The red curves in both panels are for fiducial parameters ($E_0=1\times10^{49}$~erg, $\gamma_0=10$, 
$p=2.3$, $n_0=1$~cm$^{-3}$, $\epsilon_e=0.1$, $\epsilon_B=0.01$, and $z=0.09$).
(Left panel)
Green, blue, and purple curves are for $p=2.1$, $p=2.5$, and $\gamma_0=100$
with the other parameters fixed at their fiducial values.
(Right panel)
Green, blue, and purple curves are for $n_0=0.001$, 0.01, and 0.1~cm$^{-3}$
with the other parameters fixed at their fiducial values.}
  \end{center}
 \label{fig1}
\end{figure}

\begin{figure}[t]
  \begin{center}
        \includegraphics[width=70mm]{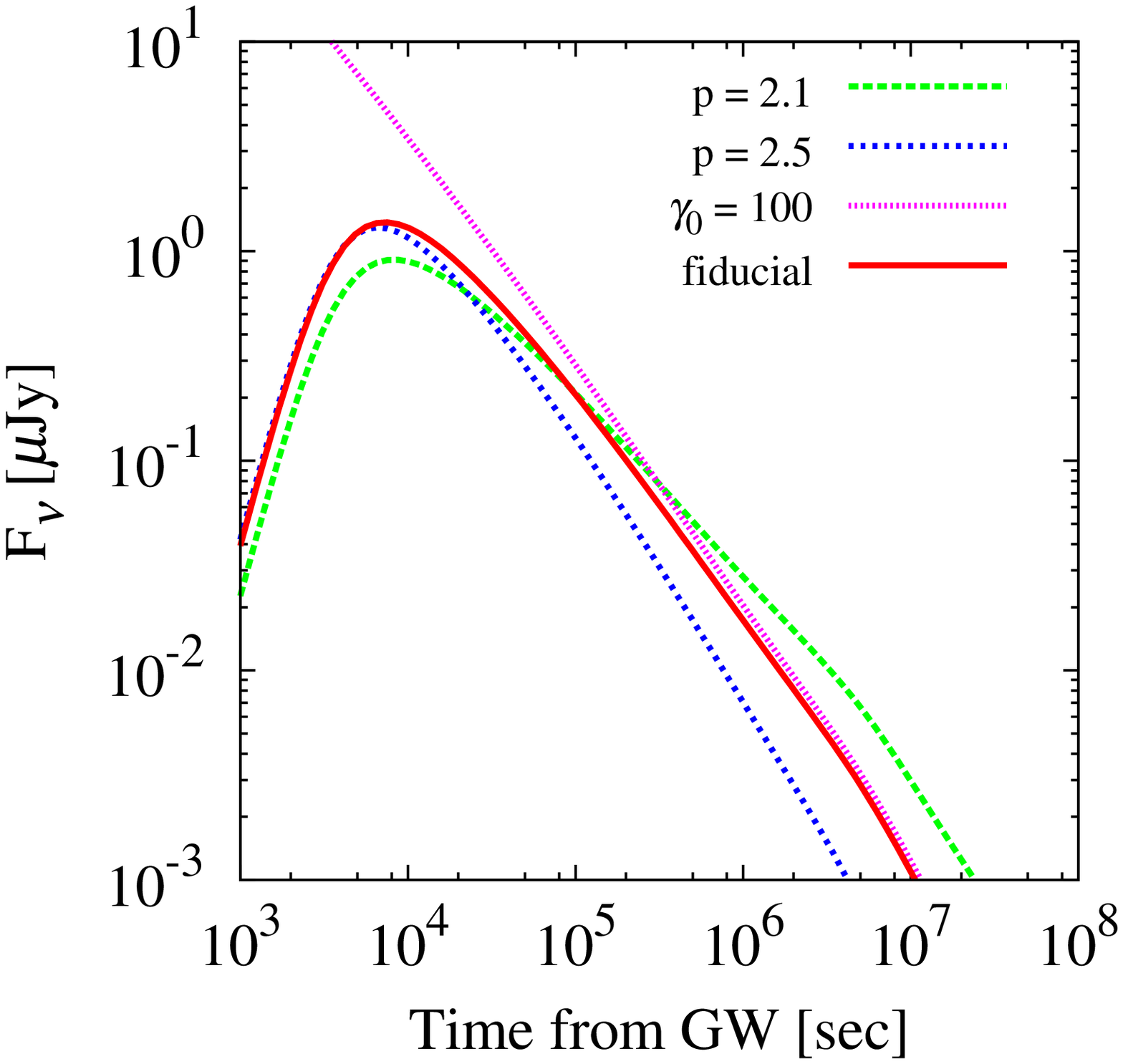}
                \includegraphics[width=70mm]{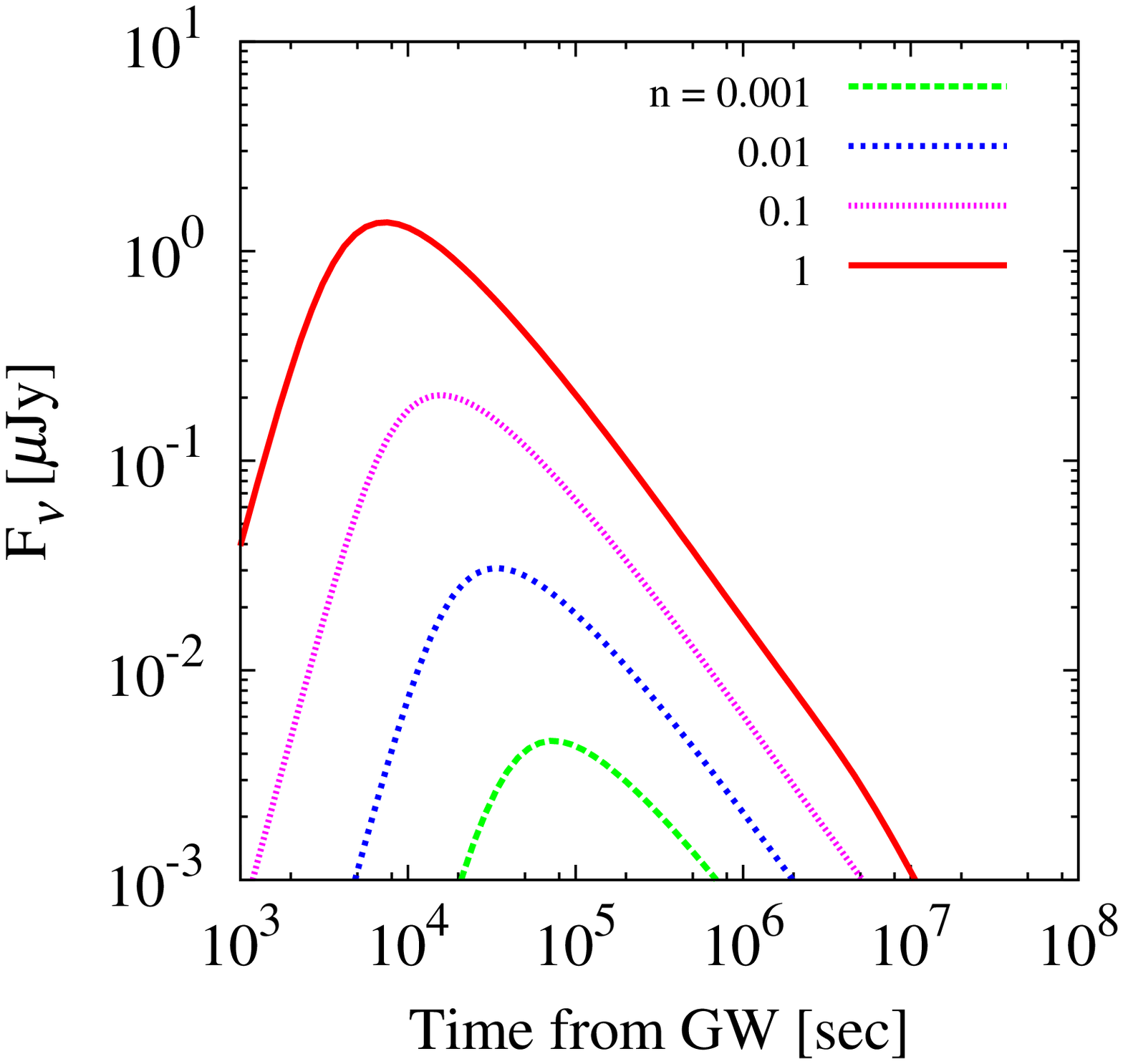} 
    \vspace{-1.3 cm}
    \caption{Optical afterglow light curves in the R-band.
The red curves in both panels are for fiducial parameters ($E_0=1\times10^{49}$~erg, $\gamma_0=10$, 
$p=2.3$, $n_0=1$~cm$^{-3}$, $\epsilon_e=0.1$, $\epsilon_B=0.01$, and $z=0.09$).
(Left panel)
Green, blue, and purple curves are for $p=2.1$, $p=2.5$, and $\gamma_0=100$
with the other parameters fixed at their fiducial values.
(Right panel)
Green, blue, and purple curves are for $n_0=0.001$, 0.01, and 0.1~cm$^{-3}$
with the other parameters fixed at their fiducial values.}
  \end{center}
 \label{fig2}
\end{figure}

\section{Discussions}

The possible gamma-ray detection with Fermi 
may help localize the GW signal, while the localization with only
GW detectors is pretty wide to follow up with electromagnetic telescopes
\cite{Fairhurst2011}.
Similarly to the case of binary neutron star merger, binary BH merger events can be
followed by radio telescopes with sensitivity $\sim10~\mu$Jy, such as 
the Very Large Array \cite[e.g.,][]{Abbott2016_followup}.
As we have shown, the predicted radio flux of the afterglow at the time
when the GW detection was published (150~days after the event) is too dim to be
detected as a transient radio object.
However, the event GW150914 encourages us to try follow-up
with radio telescopes in future GW detections, even if
the signal shows evidence of binary BH merger.

If the radio survey for the entire error region is possible,
the observations will be able to constrain the number density of the
ambient matter (see the right panel of Figure~1).
In the case of GW150914,
if the radio afterglow is not detected at $10^5$--$10^6$~sec after the GW event,
the ambient density $n_0$ may be smaller than $\sim10^{-2}$~cm$^{-3}$.

Optical/infrared upper limits have also been given 
\citep{Abbott2016_followup,Annis2016,Kasliwal2016,Santos2016,Smartt2016}.
In these cases, the
typical survey depth is $\sim20$--22~mag, corresponding to $\sim10$--30~$\mu$Jy,
which is well above the prediction of our afterglow model.
Hence the reported optical/infrared upper limits at $t\ge10^5$~sec do not constrain 
our model parameters.
If the initial Lorentz factor $\gamma_0$ is very high, $\gamma_0\ge10^2$, then
the optical/infrared afterglow in the very early  epoch ($t\ll10^4$~sec) is brighter than
$10~\mu$Jy (see the left panel of Fig.~2), 
so that early-time ($t\ll10^4$~sec) upper limits would have given the upper
bound of  $\gamma_0$. 
Note that for $n_0\sim1$~cm$^{-3}$, the optical afterglow was
detectable until $t\sim10^5$~sec  with large telescopes such as  the
Subaru Hyper Suprime-Cam\footnote{
http://www.naoj.org/Projects/HSC/index.html
}.

We briefly comment on the X-ray flux of the afterglow.
Our model predicts that the X-ray flux peaks at $t\sim t_{\rm dec}$
and the peak flux is below 
$3\times10^{-14}$~erg~s$^{-1}$cm$^{-2}$ for fiducial parameters.
{\it XMM-Newton} performed slew survey observations 2 hours and 2 weeks
after the GW event \citep{Troja2016}, and the
{\it Swift} XRT started follow-up observations about 2 days after the GW trigger
\citep{Abbott2016_followup,Evans2016}.
However, no X-ray counterpart was detected. 
The flux upper limit is $6\times10^{-13}$--$10^{-11}$~erg~s$^{-1}$cm$^{-2}$.
The model prediction is then well below this upper limit, so that
 our afterglow model is not constrained.

In this paper, we have considered spherically symmetric outflow.
Data analysis of GW showed that inclination angle, which is the angle between the
total angular momentum and the line of sight, may be large, 
although the uncertainty is large \cite{LIGO2016a}.
Hence, it seems unnatural that a collimated outflow is launched in the direction near the line of sight.
On the other hand, it may be expected that a collimated jet with high bulk Lorentz factor ($\gamma_0\ge10^2$)
is launched along the total angular momentum.
Such a jet causes a (short) GRB, 
if it is seen on-axis along the direction of the total angular momentum.
However, the gamma-ray emission detected with Fermi may not be the off-axis
emission from such a short GRB. As long as the angle between the jet axis 
and our line of sight is large,
the observed typical photon energy
(that is, the peak photon energy in the $\nu F_\nu$ spectrum) becomes smaller than
that observed in the on-axis direction  due to the relativistic Doppler effect 
\cite[e.g.,][]{Ioka2001,Yamazaki2002,Yamazaki2004}.
The typical photon energy for an on-axis observer should be
much higher than the
actually observed value ($\sim5$~MeV), which is atypical for short GRBs 
\cite[e.g.,][]{Tsutsui2013}.
So the gamma-ray emission may be emitted quasi-isotropically in this event.

For the case of off-axis ultrarelativistic jet emission, 
the isotropic-equivalent kinetic energy of the jet
for an on-axis observer
may be much higher than $10^{49}$~erg as seen in the usual short GRBs.
If the collimation-corrected kinetic energy of the jet is larger than $10^{49}$~erg,
late-time ($t\ge10^6$~sec) radio afterglow emission is brighter 
than our present result for isotropic outflow with $E_0=1\times10^{49}$~erg.
This is because all the emission energy can be seen as the jet is decelerated to a
non-relativistic regime at which the relativistic beaming effect becomes negligible
\cite[e.g.,][]{Granot2002,Granot2003}.
For example, Morsony~{\it et~al.}~\cite{Morsony2016} calculated the off-axis afterglow 
in the case of an isotropic-equivalent jet kinetic energy of $10^{51}$~erg, a jet
opening half-angle of 10~degree (corresponding to a collimation-corrected kinetic 
energy of the jet of $\sim10^{50}$~erg), and a viewing angle of the jet of 30~degree,
and they obtained the 1.4~GHz peak flux of 30~mJy for $n_0=1$~cm$^{-3}$ at  time
$\sim3\times10^6$~sec.
Hence, the presence of the relativistic jet may be tested by late-time radio observations.

\section*{ACKNOWLEDGMENTS}

\

We would like to thank the referee for careful reading of the manuscript.
%
This work was supported in part by
Grant-in-Aid  for Scientific Research
of the Japanese Ministry of Education, Culture, Sports, Science
and Technology, No.15K05088 (R.Y.), No. 25400227 (K.A.).



\end{document}